\documentclass[twocolumn,secnumarabic,amssymb, nobibnotes, aps, prd]{revtex4}
\usepackage{graphicx}

\begin{document}
\title{Contradiction between the results of observations of resistance and critical current quantum oscillations in asymmetric superconducting rings}
\author{V. L. Gurtovoi, S. V. Dubonos, S. V. Karpii, A. V. Nikulov, and V. A. Tulin}
\affiliation{Institute of Microelectronics Technology and High Purity Materials, Russian Academy of Sciences, 142432 Chernogolovka, Moscow District, RUSSIA.} 
\begin{abstract} Magnetic field dependences of critical current, resistance, and rectified voltage of asymmetric (half circles of different widths) and symmetrical (half circles of equal widths) aluminum rings were measured in the temperature region close to the superconducting transition. All these dependences are periodic magnetic field functions with periods corresponding to the flux quantum in the ring. The periodic dependences of critical current measured in opposite directions were found to be close to each other for symmetrical rings and shifted with respect to each other by half the flux quantum in asymmetric rings with ratios between half circle widths of from 1.25 to 2. This shift of the dependences by a quarter of the flux quantum as the ring becomes asymmetric makes critical current anisotropic, which explains the rectification effect observed for asymmetric rings. Shifts of the extrema of the periodic dependences of critical current by a quarter of the flux quantum contradict directly to the results obtained by measuring asymmetric ring resistance oscillations, whose extrema are, as for symmetrical rings, observed at magnetic fluxes equal to an integer and a half of flux quanta.
 \end{abstract}

\maketitle

\narrowtext

\section*{INTRODUCTION}
Superconductivity is a macroscopic quantum phenomenon. One of the experimental evidences of this is quantum oscillations in superconducting structures. In 1962, Little and Parks \cite{1} observed resistance oscillations for a thin-walled superconducting cylinder in a magnetic field. The period of oscillations corresponded to the magnetic flux quantum. More recently, when advances in technology allowed fairly small planar superconducting structures to be manufactured, similar oscillations were observed for superconducting rings. According to the generally accepted views \cite{2}, periodic resistance changes are caused by periodic changes in the allowed value of the velocity of superconducting pairs $v_{s}$, 
$$v_{s} = \frac{2\pi \hbar}{m}(n - \frac{\Phi }{\Phi _{0}})     \eqno{(1)}$$
corresponding to the lowest energy, that is, to the smallest $(n - \Phi /\Phi _{0})^{2}$ value. Here, $m$ is the mass of the superconducting pair, $n$ is an integer that determines the quantized value of pair momentum circulation along ring circles or cylinder section of length $l$, $\Phi $ is the total magnetic flux inside circle $l$, and $\Phi _{0} = \pi \hbar /e$ is the flux quantum. The amplitude of resistance oscillations $R(\Phi /\Phi _{0})$ is proportional to the square of the ratio between the correlation length $\xi (0)$ of the superconductor and the radius $r$ of the ring \cite{2}. For this reason, Little-Parks oscillations are largely observed for aluminum \cite{3} characterized by the largest correlation length.

In \cite{4}, the effect of alternating current rectification was observed for asymmetric aluminum rings at tem-peratures close to the superconducting transition tem-perature, $T = (0.950 \div 0.995)T_{c}$ The sign and magnitude of rectified voltage $V_{dc}(\Phi /\Phi _{0})$ periodically changed in a magnetic field with a period corresponding to the flux quantum $\Phi _{0} $ inside the ring. Obviously, this effect, like Little-Parks resistance $R(\Phi /\Phi _{0})$ oscillations, was caused by the quantization of velocity (1) of superconducting pairs in the ring. The periodic dependences of rectified voltage $V_{dc}(\Phi /\Phi _{0})$ always intersected zero at magnetic flux values multiple to the flux quantum, $\Phi = n\Phi _{0}$ and half the flux quantum, $\Phi = (n+0.5)\Phi _{0}$ This corresponded to the expected dependence of the mean thermodynamic value of the velocity of superconducting pairs in the ring, $\overline{v_{s}} \propto \overline{n} - \Phi /\Phi _{0}$.

Because the energy difference between the allowed velocity (1) values in a real superconducting ring is much larger than the energy of thermal fluctuations \cite{5}, the mean quantum number value $\overline{n}$ is close to the integer value $\overline{n} \approx n$ corresponding to the allowed state with minimum energy, that is, the minimum $(n - \Phi /\Phi _{0})^{2}$ value. An exception is magnetic fluxes close to $\Phi = (n+0.5)\Phi _{0}$, at which two states with opposite velocity directions have equal energies,  
$$(n - \frac{\Phi}{\Phi _{0}})^{2} = (-0.5)^{2} = (n+1 - \frac{\Phi}{\Phi _{0}})^{2} = (0.5)^{2}  $$
For this reason, $\overline{v_{s}} \propto \overline{n} - \Phi /\Phi _{0}$ is zero at both $\Phi = n\Phi _{0}$ and $\Phi = (n+0.5)\Phi _{0}$ although the state with zero velocity is, according to (1), forbidden at $\Phi = (n+0.5)\Phi _{0}$ 

\begin{figure}[b]
\includegraphics{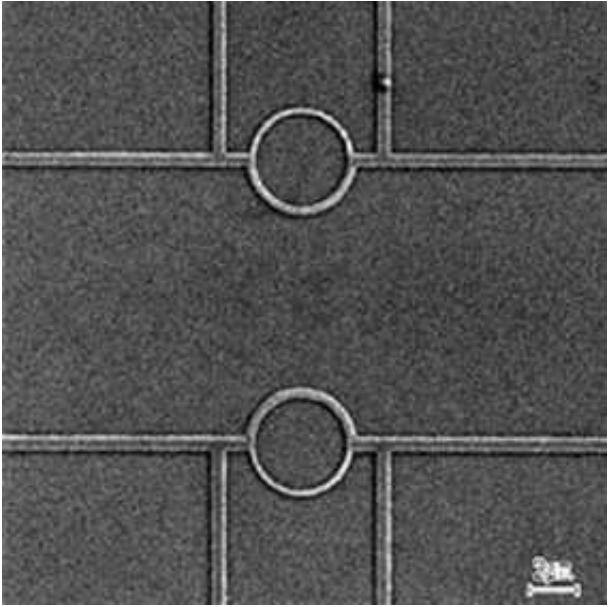}
\caption{\label{fig:epsart} Example of microstructures used in measurements. Two asymmetric rings with current and voltage contacts are shown. The widths of the half circles of the upper ring are $w_{n} = 0.2 \ \mu m$ and $w_{w} = 0.3 \ \mu m$, and those of the lower ring, $w_{w} = 0.35 \ \mu m$ and $w_{n} = 0.2 \ \mu m$.}
\end{figure}

The observation of rectified voltage $V_{dc}(\Phi /\Phi _{0})$ quantum oscillations \cite{6} presupposes that the current-voltage characteristics of an asymmetric ring should be asymmetric at $\Phi \neq n\Phi _{0}$ and $\Phi \neq (n+0.5)\Phi _{0}$ and that this asymmetry should change periodically in a magnetic field. Our measurements showed that this asymmetry was caused by critical current anisotropy, that is, the difference of the $I_{c+}$ and $I_{c-}$ values measured in opposite directions at which the transition to the resistive state is observed. Periodic changes in the sign and magnitude of critical current anisotropy 
$$I_{c,an}(\frac{\Phi}{\Phi _{0}}) = I_{c+}(\frac{\Phi}{\Phi _{0}}) - I_{c-}(\frac{\Phi}{\Phi _{0}})  \eqno{(2)}$$
explain periodic changes in rectified voltage. We found that the periodic magnetic field dependences of the critical current measured in opposite directions are close in shape but shifted by $\Delta \phi $ with respect to each other,
$$I_{c-}(\frac{\Phi}{\Phi _{0}}) \approx  I_{c+}(\frac{\Phi}{\Phi _{0}}+  \Delta \phi )  \eqno{(3)}$$
The critical current anisotropy equal  
$$I_{c,an}(\frac{\Phi}{\Phi _{0}})   \approx  I_{c+}(\frac{\Phi}{\Phi _{0}}) - I_{c+}(\frac{\Phi}{\Phi _{0}}+  \Delta \phi ) \neq 0  \eqno{(4)}$$
according to (2) and (3) is largely caused by the shift $\Delta \phi $.

In this work we present the results obtained in studies of the dependence of the shift $\Delta \phi $ on ring anisotropy. For this purpose, we measured quantum oscillations of the critical currents $I_{c+}(\Phi /\Phi _{0})$ and $I_{c-}(\Phi /\Phi _{0})$ in rings with different asymmetries, that is, with different ratios between half circle widths, and in symmetrical rings with equal half circle widths. These rings were also used to measure resistance $R(\Phi /\Phi _{0})$ and rectified voltage $V_{dc}(\Phi /\Phi _{0})$ oscillations. 

\section{EXPERIMENTAL}  
We used film aluminum nanostructures with the superconducting transition temperature $T_{c} \approx  1.23 \div 1.27 \ K$, resistance per square of approximately $0.5 \Omega /\diamond $ at 4.2 K, and the resistance ratio  $R(300 \ K)/R(4.2 \ K) \approx 3$. Measurements were taken over the temperature range from $0.95T_{c}$ to $T_{c}$. The diameter of all rings was $2r \approx  4 \ \mu m$. We studied asymmetric rings with half circle widths (Fig. 1): $w_{w} \approx  0.4 \ \mu m$, $w_{n} \approx  0.2 \ \mu m$, $w_{w}/w_{n} \approx 2$; $w_{w} \approx  0.35 \ \mu m$, $w_{n} \approx  0.2 \ \mu m$, $w_{w}/w_{n} \approx 1.75$; $w_{w} \approx  0.3 \ \mu m$, $w_{n} \approx  0.2 \ \mu m$, $w_{w}/w_{n} \approx 1.5$; $w_{w} \approx  0.25 \ \mu m$, $w_{n} \approx  0.2 \ \mu m$, $w_{w}/w_{n} \approx 1.25$ and symmetrical rings with $w_{w} \approx  0.4 \ \mu m$, $w_{n} \approx  0.4 \ \mu m$, $w_{w}/w_{n} \approx 1$. The structures were prepared from films of thickness $d = 40 \div 50 \ nm$. The half circle section areas were $s_{w} = dw_{w} \approx  0.01 \div 0.02 \ \mu m^{2}$; $s_{n} = dw_{n} \approx  0.008 \div 0.02 \ \mu m^{2}$. The microstructures were prepared on silicon sub-strates using a JEOL-840A electron microscope, which was transformed into a laboratory electron lithograph with the use of the NANOMAKER package.

Measurements were taken by the four-probe method (Fig. 1) in a glass helium cryostat using $^{4}He$ as a coolant. The temperature could be decreased to $1.19 \ K$ by pumping helium off. The temperature was measured by a calibrated thermal resistance ($R(300 \ K) = 1.5 \ k\Omega $) with a $0.1 \ \mu A$ measuring current. Magnetic field {\bf B} perpendicular to the plane of the sample was created by a copper solenoid situated outside the cryostat. The dependences of values on the current $I_{sol}$ through the solenoid were recorded. The magnetic field created by a current in the solenoid was found from the calibration plot $B_{sol} = k_{sol}I_{sol}$ with the use of a Hall probe ($k_{sol} = 129 \ G/A$). The period $B_{0} = \Phi _{0}/S = 1.4 \div 1.6 \ G$ of all periodic dependences $R(B)$, $V_{dc}(B)$, and $I_{c}(B)$ corresponded to the area $S = \pi r^{2} = 14.8 \div 13.0 \ \mu m^{2}$ of the ring used for measurements; here, $r = 2.2 \div 2.0 \ \mu m$ is a value close to the inside ring radius. To decrease the magnetic field of the Earth, the cryostat region with the sample was screened by a permalloy cylinder. The residual magnetic field was $B_{res} \approx  0.15 \ G$, that is, approximately one tenth of the $B_{0}$ period.

The magnetic field dependences of the critical currents $I_{c+}(B)$ and $I_{c-}(B)$ were determined by measuring periodically repeating current-voltage characteristics (a period of 10 Hz) in a slowly varying magnetic field $B_{sol}$ (a period of approximately 0.01 Hz) as follows. First, the condition that the structure was in the superconducting state was checked. Next, after the threshold voltage was exceeded (this voltage, set above induced voltages and noises of the measuring system, determined the minimum measurable critical current), magnetic field and critical current (with a delay of about $30 \ \mu s$) were switched on. This procedure allowed us to measure sequentially critical currents in the positive $I_{c+}$  and negative $I_{c-}$  directions with respect to the external measuring current $I_{ext}$. Measurements of one $I_{c+}(B)$ or $I_{c-}(B)$  dependence (1000 values) took about 100 s. Little-Parks oscillations $R(B) = V(B)/I_{ext}$ were recorded at a constant $I_{ext} = 0.1 \div 2.0 \ \mu A$ current. The field dependences of rectified voltage $V_{dc}(B)$ were measured using sinusoidal current $I_{ext}(t) = I_{0}\sin (2\pi ft)$ with the amplitude $I_{0}$ up to $50 \ \mu A$ and frequency $f = 0.5 \div 5 \ kHz$. Because of incomplete screening, the minima of the $R(B_{sol})$ solenoid field dependences of resistance and zero rectified voltage $V_{dc}(B_{sol})$ were shifted by $-B_{res} \approx  -0.15 \ G$. The $R(B_{sol} + B_{res})$ dependences had minima at $B_{sol} + B_{res} = n\Phi _{0}/S$ and maxima at $B_{sol} + B_{res} = (n+0.5)\Phi _{0}/S$ and the $V_{dc}(B_{sol} + B_{res})$ dependences intersected zero at these values of the total magnetic field $B_{sol} + B_{res}$. Because the simultaneous reversal of the total external field {\bf B} and measuring current $I_{ext}$ was equivalent to rotation through $180^{o}$, the equality $I_{c+}(B) = I_{c-}(-B)$ had to be satisfied. We had $I_{c+}(B_{sol} + B_{res}) = I_{c-}(-B_{sol} - B_{res})$ for all the dependences measured. This proves that $B_{sol} + B_{res}$ was the total external field also in critical current measurements.

\begin{figure}
\includegraphics{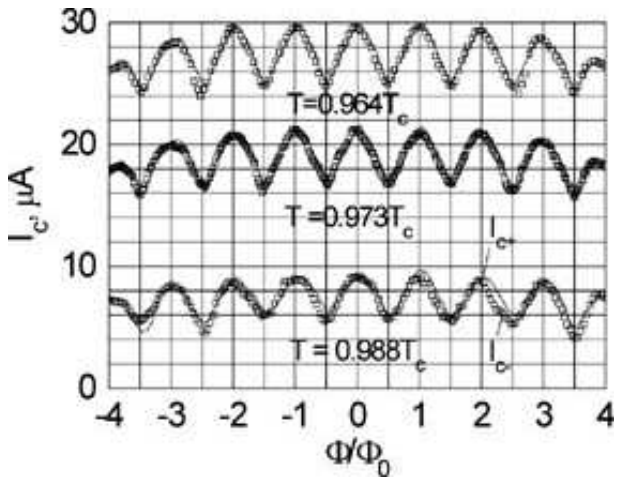}
\caption{\label{fig:epsart} Magnetic field dependences of the critical current of a symmetrical ring of width $w_{w} \approx  w_{n} \approx  0.4 \ \mu m$ measured in the positive $I_{c+}$ (solid lines) and negative $I_{c-}$ (squares) directions at $T= 0.964T_{c}$, $0.973T_{c}$ and  $0.988T_{c}$.}
\end{figure}

Generally, the total flux in the ring is
$$\Phi  = \Phi _{ext}  + \Phi _{Iext} + \Phi _{Ip} \eqno{(5)}$$
where $\Phi _{ext} = S(B_{sol} + B_{res})$  is the flux created by external fields and
$$\Phi _{Iext} + \Phi _{Ip} = LI_{ext}\frac{s_{w} - s_{n}}{2(s_{w} + s_{n})} + LI_{p} \eqno{(6)}$$
are the fluxes created by the measuring current $I_{ext}$ and the persistent current $I_{p}$, which exists by virtue of quantization condition (1). At ring inductances $L \approx  2 \times 10^{-11} \ H$ and maximum measuring current $I_{ext} = 30 \ \mu A$ used in this work and the persistent current over the temperature range studied, additional fluxes $\Phi _{Iext}$ and $\Phi _{Ip}$ did not exceed several hundredths of the flux quantum $\Phi _{0}$. We therefore used the approximation
$$\Phi _{ext} = \Phi _{ext}  + \Phi _{Iext} + \Phi _{Ip} \approx \Phi _{ext} = S(B_{sol} + B_{res})$$
The dependences measured are represented as functions of the magnetic flux inside rings,
$$\frac{\Phi }{\Phi _{0}} = \frac{SB}{\Phi _{0}} = \frac{S(B_{sol} + B_{res})}{\Phi _{0}}$$
created by external fields $B_{sol} + B_{res}$. The exact $S$ value was selected using the condition that the periods of oscillations of the $I_{c}(\Phi )$, $R(\Phi )$ and $V_{dc}(\Phi )$ values should equal the flux quantum $\Phi _{0}$.

\begin{figure}
\includegraphics{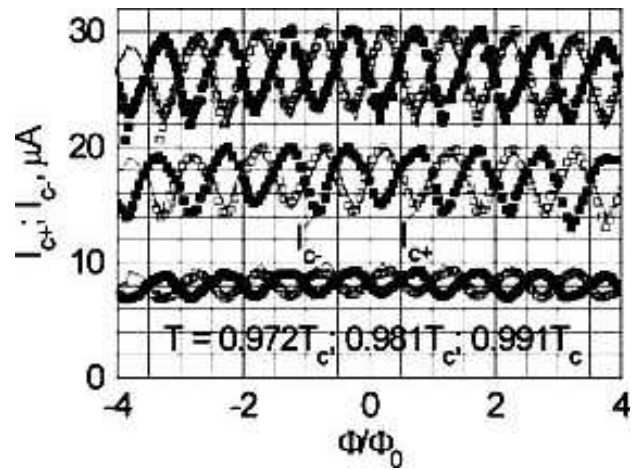}
\caption{\label{fig:epsart} Magnetic field dependences of the critical current of an asymmetric ring with half circles of widths $w_{w} = 0.25 \ \mu m$ and $w_{n} = 0.2 \ \mu m$ measured in the positive $I_{c+}$ (solid lines) and negative $I_{c-}$ (black squares) directions at $T = 0.972T_{c}$,   $0.981T_{c}$, and $0.991T_{c}$. The shifted dependence $I_{c-}(\Phi /\Phi _{0} + 0.5)$ is also shown (light squares).}
\end{figure}

\section{RESULTS AND DISCUSSION}
Clearly, the magnitudes and magnetic field depen-dences of the critical currents measured in opposite directions for an absolutely symmetrical ring should coincide $I_{c+}(\Phi /\Phi _{0}) = I_{c-}(\Phi /\Phi _{0})$, because rotation through $180^{o}$ about the magnetic field direction should not then cause any changes. Our measurements showed that such dependences measured for rings with half circles of equal widths $w_{w} \approx w_{n} \approx  0.4 \ \mu m$, $w_{w}/w_{n} \approx 1$ were close to each other (Fig. 2), 
$$I_{c+}(\Phi /\Phi _{0}) \approx I_{c-}(\Phi /\Phi _{0})$$ 
The closest coincidence was, however, observed when one of the dependences was shifted by $\Delta \phi = 0.05 \pm  0.02$ of the flux quantum. The absence of the exact equality $I_{c+}(\Phi /\Phi _{0}) = I_{c-}(\Phi /\Phi _{0})$ is likely evidence that the rings used were not absolutely symmetrical. Measurements for rings with controlled asymmetry showed that even comparatively small asymmetry, $w_{w}/w_{n} \approx 1.25$ (half circles of widths $w_{w} \approx  0.25 \ \mu m$ and $w_{n} \approx  0.2 \ \mu m$) resulted in a maximum relative shift $\Delta \phi = 0.5 \pm  0.02$ of the dependences of the critical currents measured in the opposite directions (Fig. 3),  
$$I_{c+}(\Phi /\Phi _{0}) \approx I_{c-}(\Phi /\Phi _{0} + 0.5)$$ 
This shift remained unchanged as the $s_{w}/s_{n} = w_{w}/w_{n}$ ratio of ring section areas increased from 1.25 to 2.00 (Fig. 4).

\begin{figure}
\includegraphics{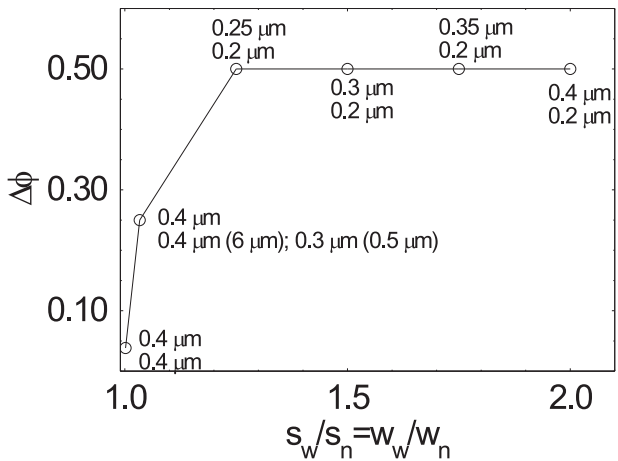}
\caption{\label{fig:epsart} Dependence of the shift $\Delta \phi $ at which the closest coincidence of critical current oscillations $I_{c+}(\Phi /\Phi _{0}) \approx I_{c-}(\Phi /\Phi _{0} + \Delta \phi )$ measured in the opposite directions is observed on ring anisotropy. At equal half circle widths ($w_{w} \approx w_{n} \approx  0.4 \ \mu m$), the shift is $\Delta \phi \approx  0.05 $ but narrowing of even a short ($0.5 \ \mu m$) portion of one of half circles with semicircumference length $\pi r \approx  6 \ \mu m$ from 0.4 to 0.3 $\mu m$ increases the shift to $\Delta \phi \approx  0.25 $. At a $w_{w}/w_{n} \approx 1.25$ ratio between half circle widths ($w_{w} \approx  0.25 \ \mu m$ and $w_{n} \approx  0.2 \ \mu m$), the shift reaches $\Delta \phi \approx  0.5 $. It remains unchanged as the $w_{w}/w_{n} $ ratio increases above 1.25.}
\end{figure}

The shift $\Delta \phi $ was also independent of the measuring current. Over the temperature range studied, the critical current changed by a factor of 10 (from 3 to 30 $\mu A$) and the closest coincidence of the $I_{c+}(\Phi /\Phi _{0}) \approx I_{c-}(\Phi /\Phi _{0} + \Delta \phi )$ dependences was always observed at $\Delta \phi = 0.50 \pm  0.02$. For absolutely symmetrical rings, $I_{c+}(\Phi /\Phi _{0}) = I_{c-}(\Phi /\Phi _{0})$, and the result obtained therefore means that the appearance of ring asymmetry shifts the dependences of the critical current (to opposite sides for measurements in the opposite directions) by $\Delta \phi /2$, or a quarter of the flux quantum at $s_{w}/s_{n} = w_{w}/w_{n} \geq 1.25$. 

This shift of the dependences is responsible for the critical current anisotropy (4) and the effect of alternating current rectification (Fig. 5) observed in asymmetric rings \cite{4}. The existence of the shift, however, contradicts both the dependence of the velocity of superconducting pairs corresponding to quantization condition (1) and the results of measurements of resistance oscillations for asymmetric rings. Clearly, the rectification effect cannot be observed in an absolutely symmetrical ring, where $I_{c+}(\Phi /\Phi _{0}) = I_{c-}(\Phi /\Phi _{0})$. All the existing concepts of quantum effects in superconductors lead us to suggest that the rectification effect observed for asymmetric rings \cite{4} is caused by a change (resulting from ring asymmetry) in the functions that describe the magnetic field dependence of the critical current. Our measurements, however, show that the equality of the critical currents measured in the opposite directions changes for their inequality because of changes in the arguments of the functions rather than the functions themselves,
$$ I_{c+}(\frac{\Phi}{\Phi _{0}}) = I_{c-}(\frac{\Phi}{\Phi _{0}})  \longrightarrow  I_{c+}(\frac{\Phi}{\Phi _{0}}+0.25) \neq I_{c-}(\frac{\Phi}{\Phi _{0}}-0.25)$$
This is a very strange result, because the periodic $I_{c+}(\Phi /\Phi _{0})$ and $I_{c-}(\Phi /\Phi _{0})$ dependences are actually functions of $\Phi /\Phi _{0} - n$.

\begin{figure}
\includegraphics{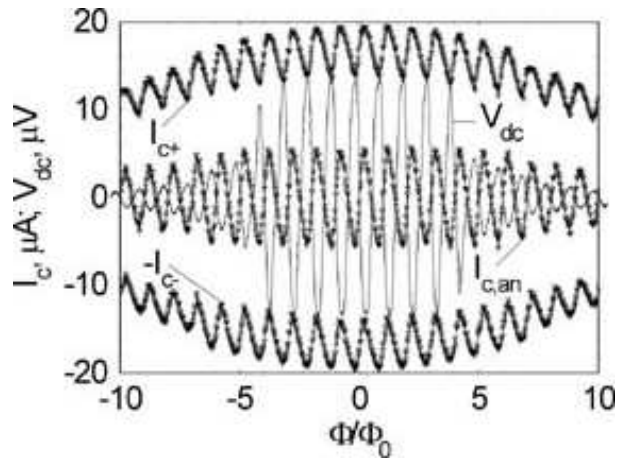}
\caption{\label{fig:epsart} Quantum oscillations in magnetic field of critical currents $I_{c+}$, $I_{c-}$, critical current anisotropy $I_{c,an} = I_{c+}-I_{c-}$, and rectified voltage $V_{dc}$ measured for an asymmetric ring ($w_{w} \approx  0.4 \ \mu m$ and $w_{n} \approx  0.2 \ \mu m$) at $T \approx  0.987T_{c}$.}
\end{figure}

The shift of the $I_{c+}(\Phi /\Phi _{0})$ and $I_{c-}(\Phi /\Phi _{0})$ dependences might have a trivial explanation similar to that valid for the extrema of the dependences of the critical current in asymmetric superconducting quantum interferometers \cite{7} if the flux $\Phi _{Iext}$ created by the measuring current in asymmetric rings, see (6), was fairly large and the shift value $\Delta \phi $ depended on the asymmetry value $(s_{w} - s_{n})/2(s_{w} + s_{n})$ and current $I_{ext}$. But the small $\Phi _{Iext}$ value and the independence of the shift value from ring asymmetry (Fig. 4) and critical current (Fig. 3) show this explanation to be inapplicable and are evidence that the total flux in the ring (5) insignificantly differs from the $\Phi _{ext} = S(B_{sol} + B_{res})$ flux created by the external magnetic field. This means that the change in the arguments of the $I_{c+}(n-\Phi /\Phi _{0})$ and $I_{c-}(n-\Phi /\Phi _{0})$ functions observed as the ring becomes asymmetric can only be related to a change in the quantum number $n$.

\begin{figure}
\includegraphics{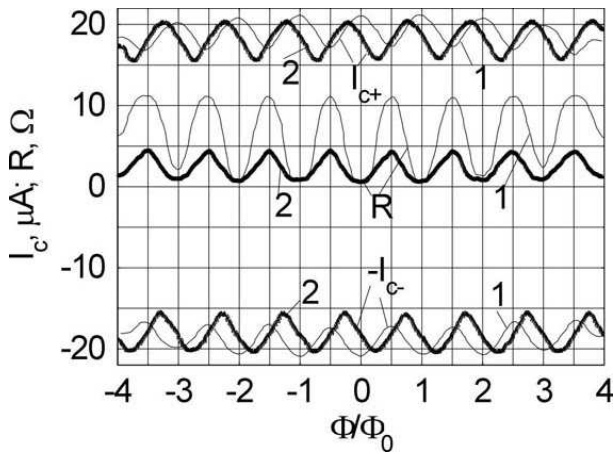}
\caption{\label{fig:epsart} Quantum oscillations in magnetic field of resistance $R$ at $T \approx  T_{c}$ and critical currents $I_{c+}$, $I_{c-}$, measured in the opposite directions at $T\approx  0.97T_{c}$ for symmetrical (1), $w_{w} \approx  w_{n} \approx  0.4 \ \mu m$, and asymmetric  (2), $w_{w} \approx  0.3 \ \mu m$ and $w_{n} \approx  0.2 \ \mu m$, rings.}
\end{figure}

The reason for the periodic character of the $I_{c+}(\Phi /\Phi _{0})$ and $I_{c-}(\Phi /\Phi _{0})$ dependences for both symmetrical and asymmetric rings can be no other than periodic changes in the velocity of superconducting pairs, whose allowed values are determined by quantization condition (1). Maxima of the dependences $I_{c+}(\Phi /\Phi _{0}) = I_{c-}(\Phi /\Phi _{0})$ recorded for symmetrical rings correspond to $\Phi = n\Phi _{0}$ and minima, to $\Phi = (n+0.5)\Phi _{0}$ that is, to zero and maximum velocity modulus values at the lowest level allowed by condition (1). The $I_{c+}(\Phi /\Phi _{0})$ and $I_{c-}(\Phi /\Phi _{0})$ dependences remain periodic in asymmetric rings, but the positions of their extrema shift by $ \pm 0.25\Phi _{0}$. Seemingly, this should mean $ \pm 0.25\Phi _{0}$ shifts in the positions of the extrema of the modulus of the velocity of superconducting pairs as the ring becomes asymmetric. This suggestion, however, contradicts the results of Little-Parks resistance oscillation measurements performed by us for the same asymmetric rings.

It is generally recognized \cite{2} that resistance oscillations are a consequence of oscillations of the square of the velocity of superconducting pairs,
$$\Delta R(\frac{\Phi }{\Phi _{0}}) \propto  v_{s}^{2}(\frac{\Phi }{\Phi _{0}}) \propto (n -\frac{\Phi }{\Phi _{0}})^{2} $$
Accordingly and in agreement with quantization condition (1), $\Delta R(\Phi /\Phi _{0})$ dependence minima and maxima are observed at $\Phi = n\Phi _{0}$ and  $\Phi = (n+0.5)\Phi _{0}$, respectively. Our measurements showed (Fig. 6) that ring asymmetry caused no shift of the $R(\Phi /\Phi _{0})$ function extrema.

To summarize, our experiments revealed a contradiction between the results of resistance oscillation $R(\Phi /\Phi _{0})$ measurements and the behavior of critical currents $I_{c+}(\Phi /\Phi _{0})$ and $I_{c-}(\Phi /\Phi _{0})$ in asymmetric superconducting rings. Ring asymmetry does not cause shifts of $R(\Phi /\Phi _{0})$ extrema for resistance oscillations, whereas, for critical current oscillations, the extrema of the dependences $I_{c+}(\Phi /\Phi _{0})$ and $I_{c-}(\Phi /\Phi _{0})$ shift by $ \pm 0.25\Phi _{0}$. This result was obtained for all the rings studied with asymmetry $w_{w}/w_{n} \geq  1.25$. This contradiction remains an open question, because resistance and critical current oscillations cannot be caused by anything other than quantization (1) of the velocity of superconducting pairs.

\section{CONCLUSIONS}
We are led to conclude that shifts of magnetic field dependences of the critical current observed in this work as superconducting ring asymmetry appears is a very strange and inexplicable phenomenon. Note that this result is absolutely reliable experimentally and can not be an artifact. Although we were unable to prepare an absolutely symmetrical ring, the difference between the shifts equal to $\Delta \phi = 0.05 \pm  0.02$ in rings with $w_{w}/w_{n} \approx 1$ and $\Delta \phi = 0.5 \pm  0.02$ in rings with $w_{w}/w_{n} \geq  1.25$ is evidence that the observed change in the arguments of the $I_{c+}(\Phi /\Phi _{0})$ and $I_{c-}(\Phi /\Phi _{0})$ functions cannot be an artifact. We also obtained a shift of an intermediate value, $\Delta \phi \approx 0.25 $ (Fig. 4). To create asymmetry smaller than $w_{w}/w_{n} =  1.25$, we prepared a ring with equal half circle widths of $\approx 0.4 \ \mu m$ along the major portion of their lengths $l_{w} = l_{n} = \pi r \approx  6 \mu m$ and a narrowing to along a short segment of length $\approx 0.5 \ \mu m$ in one half circle. Such small but controlled asymmetry gave a shift of $\Delta \phi \approx 0.25 $, which was substantially larger than the shift $\Delta \phi \approx 0.05 $ for rings with uncontrolled asymmetry. The shift $\Delta \phi = 0.05 \pm  0.02$ observed for rings prepared as symmetrical is evidence of high accuracy of measurements of the difference between $\Phi /\Phi _{0} - n$ arguments in symmetrical and asymmetric rings. Such a high accuracy is a result of both the method used to measure the $I_{c+}(\Phi /\Phi _{0})$ and $I_{c-}(\Phi /\Phi _{0})$ dependences and periodic character of these functions. Measurements of the $I_{c+}$, $I_{c-}$, $I_{c+}$, $I_{c-}$, $I_{c+}$, $I_{c-}$ ... values at successive time moments exclude the possibility of the influence of magnetic field measurement errors on the accuracy of $\Delta \phi$ values. We observed up to 30 periods of oscillations, which allowed us to determine $\Delta \phi$ with high accuracy. All this is evidence that dependence shifts as rings become asymmetric are an unambiguous experimental fact. It only remains for us to hope that this strange fact will be explained sometime.

\section*{Acknowledgement}
This work was financially supported by the Program for Basic Studies of the Division for Information Tech-nologies and Computational Systems of the Russian Academy of Sciences "Organization of Calculations with the Use of New Physical Principles" within the framework of the project "Quantum Bit on the Basis of Micro and Nanostructures with Metallic Conductivity," the program of the Presidium of the Russian Academy of Sciences "Quantum Nanostructures," and the Russian Foundation for Basic Research (project no. 04-02-17068).

\end{document}